\documentclass[12pt,preprint]{aastex}
\usepackage{natbib}
\usepackage{emulateapj5}  

\def\ergcm2s{erg cm$^{-2}$ s$^{-1}$} 
\def\ergs{erg s$^{-1}$}		
\def\etal{et al.}		
\def\n4038{NGC 4038/39}		
\def\chandra{{\it Chandra}}
          
\def\ergcm2s{~erg cm$^{-2}$ s$^{-1}$ } 
\def\ergs{~erg s$^{-1}$}                
\def\cmsq{~cm$^{-2}$ }          %
\def\nh{~$\rm{N_{H}}$}

\def\etal{et al.~}              

\def\n4038{~NGC4038/39}         
\def\chandra{{\it Chandra }}

\def\x2{$\chi^{2}$}     

\begin{document}

\title{The Multi-Colored Hot Interstellar Medium of ``The Antennae'' Galaxies 
(NGC~4038/39)}

\author{G. Fabbiano, M. Krauss, A. Zezas, A. Rots}
\affil{Harvard-Smithsonian Center for Astrophysics,\\
60 Garden Street, Cambridge, MA 02138\\}
\author{and S. Neff}
\affil{Laboratory for Astronomy and Solar Physics,\\ 
    NASA / Goddard Space Flight Center, Greenbelt, MD 20771\\}

\bigskip

\begin{abstract}
We report the results of the analysis of the extended soft emission
discovered in the \chandra\ ACIS pointing at the
merging system \n4038\ (the Antennae). We present a `multi-color'
X-ray image that suggests both extensive absorption by the dust in this
system, peaking in the contact region, as well as variations in the
temperature of different emitting regions of the hot interstellar
medium (ISM). Spectral fits to multi-component thermal emission models confirm this
picture and give a first evaluation of the parameters of the hot plasma.
We compare the diffuse X-ray emission with radio continuum (6cm),
HI, CO, and H$\alpha$ images to take a first look at the
multi-phase ISM of the Antennae galaxies.
We find that the hot (X-ray) and cold (CO) gas have comparable
thermal pressures in the two nuclear regions. We also conclude that
the displacement between the peak of the diffuse X-ray emission in
the north of the galaxy system, towards the inner regions of the northern
spiral arm (as defined by  H$\alpha$, radio continuum and HI), could
result from ram pressure of infalling HI clouds.
\end{abstract}


\section{Introduction}

This is the fifth in a series of papers based on the first 
\chandra\ ACIS-S (Garmire 1997; Weisskopf et al 2000) 
observation of the Antennae galaxies (\n4038).
The first paper in this series gives a general picture of the 
overall results of this 72.2~ks. observation (ObsID 315: Fabbiano et al 2001 --
Paper~I).
The following three papers (Zezas et al 2002a, b -- Paper~II, III; 
Zezas \& Fabbiano 2002 -- Paper~IV) 
discuss the properties of the point-like source population detected
in this merging system. Here, we concentrate on the analysis of
the complex diffuse soft emission reported in Paper~I
and on its importance
for our understanding of the hot ISM of these galaxies.

The Antennae (dynamically modeled by Toomre \& Toomre 1972 and Barnes
1988) are the nearest pair of colliding galaxies
involved in a major merger (D = 19~Mpc for H$_o = 75~\rm km~s^{-1}~Mpc^{-1}$
\footnote{We have adopted this distance for conformity with most recent
work on the Antennae; in Papers~I-III, a distance of 29~Mpc, H$_o = 50~\rm km~s^{-1}~Mpc^{-1}$
was used; in Paper~IV we discuss our results using both D = 19 and 29~Mpc}).
Hence, this system provides a unique opportunity for getting the most
detailed look possible at the consequences of a galaxy merger as
evidenced by induced star formation and the conditions in the ISM.
Each of the two colliding disks shows rings of giant H~II regions and bright
stellar knots with luminosities up to $M_V \sim -16$ (Rubin et al 1970),
which are resolved with  the  {\it Hubble Space Telescope}
into typically about a dozen young star clusters
(Whitmore \& Schweizer 1995, Whitmore et al 1999). These knots coincide with the peaks of
H$\alpha$, 2.2$\mu$, and 6-cm radio-continuum emission  
(Amram et al 1992; Stanford et al 1990; Neff \& Ulvestad 2000),
indicating an intensity of star formation in each single region exceeding
that observed in 30 Doradus.  CO aperture synthesis maps reveal major
concentrations of molecular gas, including $\sim$2.7$\times 10^9
M_{\odot}$ in the region where the two disks overlap
(Stanford et al 1990; Wilson et al 2000). 
A recent K-band study derives ages for the star clusters ranging from 4 
to 13~Myr, and measures high values of extinction ($A_V \sim 0.7 - 4.3$~mag;
Mengel et al 2000).

The presence of an abundant hot ISM in the Antennae was originally suggested by
the first {\it Einstein} observations of this system (Fabbiano \&
Trinchieri 1983), and has been revisited with all 
following major X-ray telescopes (ROSAT: 
Read et al 1995, Fabbiano et al 1997; and ASCA: Sansom et al 1996).
The ROSAT HRI image was used in a recent multiwavelength study of the
Antennae by Zhang et al (2001), which suggests feedback between 
star clusters and the interstellar medium (ISM).
The {\it Chandra} observation gives us the means to
study this hot ISM in unprecedented detail.
The  spatial resolution of \chandra\ is at least
10 times superior to that of any other X-ray observatory, allowing us to resolve
the emission on physical sizes of $\sim 47$~pc for D$=$19~Mpc.
With this resolution we have detected individual X-ray sources 
down to $L_X \sim 10^{38} \rm ergs~s^{-1}$ (Paper~II)
and we can image in detail the spatial properties
of more extended emission regions (see Paper~I). At the same time, ACIS
allows us to study for the first time the X-ray spectral properties of these 
emission regions, providing additional important constraints on their nature.

Our initial look at these {\it Chandra} data (Paper~I) revealed a complex, 
diffuse and soft emission component responsible for about half of the 
detected X-ray photons from the two merging galaxies. The X-ray spectrum
of this component exhibits emission lines, pointing to 
optically thin gaseous emission. In the present paper, we
explore further the spectral/spatial properties
of this emission and its relation to the multi-phase ISM:
in \S2 we present a `true-color' X-ray
image of the extended emission (i.e., an image where the color scale
represents the photon energy), in \S3 we present a spectral fitting analysis
of these data, and in \S4 we describe new 4 and 6cm radio images derived from
archival radio data, matched to the effective resolution of the diffuse
X-ray emission. In \S5  we discuss the physical properties of the
hot ISM and  we compare the 
diffuse X-ray emission with H$\alpha$, radio continuum (6cm), HI and CO
data.

\section{The True-Color X-ray Image}

\chandra\ was pointed to \n4038\ on December 1, 1999, for 72.2~ks with
the back-illuminated ACIS-S3 CCD chip at the focus.
The observations and data reduction and screening are described in
Paper~I. We used CIAO 2.2.1 and other software developed
for the purpose in the present data analysis.

We have used these data to produce a true-color X-ray image of
the Antennae that would be sensitive to spectral differences in the
soft diffuse emission. The true-color X-ray image of Paper~I showed that
the diffuse emission is soft (mostly at energies below 2~keV) and
thus differentiated it from the harder broad-band emission of the 
point-like sources, but did not have the resolution to show spectral
differences in this soft emission. To obtain this resolution,
we cut the data into
energy bands of 0.3-0.65, 0.65-1.5, and 1.5-6 keV.  These bands were
chosen using thermal plasma models as a reference: the low energy band
does not include the Fe-L blend seen around 1 keV, whereas the
high energy band should include only the hottest (or most highly
absorbed) material.  
The resulting images
were then adaptively smoothed using the tool {\it csmooth} with a minimal
significance (S/N) of
3.5 and maximum significance (S/N) of 5.0. This tool uses different 
smoothing scales to enhance low-surface-brightness extended features,
while retaining the resolution allowed by the areas of the image with
the highest signal-to-noise ratios.

Monochromatic exposure maps were created for each of the 
bands at 0.5, 1, and 3 keV,  and smoothed using the same smoothing scales as the images.  
The images were then divided by the corresponding exposure map and 
combined using {\it dmimg2jpg} to create a "true-color" image,
with a red to blue progression to represent 
increasing photon energy (Fig.~1).

To derive a true-color image of the diffuse emission, we then proceeded
to identify and remove point-like sources.
A source list was first generated using the {\it wavdetect} tool, using
scales of 1, 1.4, 2, 2.8 and 4 pixels; the initial list was then culled
to include only obvious point sources.  These are sources 
1, 2, 3, 4, 11, 12, 13, 14, 16, 18, 19, 20, 21, 23, 29, 31, 32, 33,
34, 35, 37, 38, 41, 42, 43, 44, 46, 47, 49, from the list of Paper~II.
Background regions were created to
best represent the local diffuse emission.  
The tool {\it dmfilth} 
(dmFillTheHole) was run in each of the three spectral  band images to fill in the source
regions with pixel values sampled from the Poisson distribution whose
mean was that of pixel values in the corresponding background region.  
The same procedure described above of smoothing, exposure correction and combining of the
three resulting images into a true-color image was then followed.
The result is shown in fig.~2.

While figs.~1, 2 do not give a quantitative assessment of the spectral properties, 
because the color scale is not linear and has been enhanced to show faint features,
it is clear from them that there are spectral
variations throughout. In particular, the area of impact
between the two galaxies [centered approximately on RA(J2000)=$12^h 01^m 55^s$
and Dec(J2000)=$-18^o 52^{'} 50^{''}$] releases only high energy photons, as it would
be expected in the presence of large amounts of obscuring material.
The nuclear regions, as already suggested in Paper~I,
include harder emission (`white' in figs~1, 2) than other relatively unobscured
regions of the ISM.

\section{Spectral Analysis of the Diffuse X-ray Emission}

To derive quantitative information on the spectral differences 
suggested by the color maps, we performed a spectral analysis of the diffuse
emission. 
We screened the data for all the point-like sources reported in Paper~II.
We retained the sources with a strong
extended component. These are  sources 5, 6, 7, 10, 25 of
Paper~II. Fig.~3a shows a raw image of the
Antennae together with the excised  source
regions. We did not correct
for the amount of diffuse emission which was also excised together
with the excluded sources. This does not affect the parameters of the
fit (temperature, \nh, photon index, abundance) but results in  lower normalizations
(or emission measure in the case of a thermal model).
  
To derive quantitative information on the varying spectral properties
suggested by the color image, we first divided the area in a grid, so to perform
a systematic spectral analysis from which we could derive a first-cut
spectral map of the hot ISM. In order to have at least 100 in each 
extracted spectrum (before background removal), 
we used regions $50\times50$~pixels ($24.5''\times24.5''$) in size. 
Using fig.~2 as a guide, we then focused on the regions with the 
largest number of counts (highest signal to noise) to perform a more in depth analysis.
Both grid and individual regions (ellipses) are shown superimposed on the
color map of the diffuse emission in fig.~3b.

 For each region (either square in the grid, or ellipse) 
 we extracted photon invariant (PI) spectra and created weighted response
matrices (RMFs)  and ancillary response files (ARFs) using the
$\textit{acisspec}$ script of CIAO v2.1. This script invokes
the $\textit{dmextract}$ tool to extract the spectrum and then the
$\textit{mkwrmf}$ and $\textit{mkwarf}$ tools in order to create area
weighted RMF and ARF files for each region, since they span more than
one Fits Embedded Function (FEF) regions. The spectra were then grouped
 to have at least 15 counts per bin in order to use \x2 statistic for
the fits.   From the spectrum of the diffuse emission presented in
Paper~I, is clear that most photons are
below 5~keV. Therefore in order to minimize contamination by the
background which dominates above 7~keV we ignored any data outside the
 0.3-7.0~keV energy range. For the spectral fits we used the
XSPEC~v10.0 spectral fitting package (Arnaud \etal 1996). All the
cited errors for the spectral parameters are for the 90\% confidence
level for one interesting parameter, unless otherwise stated.  The
background was obtained from source free regions outside the galaxy.
  
\subsection{Spectral `Map'}

 First we attempted fits with an absorbed  single thermal plasma
model (RS model; Raymond \& Smith 1977) and solar abundances (the abundance table
by Anders \& Grevesse 1989 was used).  For these fits the absorbing column density
was  free to vary above  the Galactic N$_H = 3.4\times10^{20}$~\cmsq 
(Stark \etal, 1992).  
Typical best-fit kT is betweeen 0.1~keV
and 0.9~keV and best-fit  \nh\ can reach as high as
$\sim8 \times 10^{21}$~\cmsq.  However, these fits were unacceptable in all but 6
cases (reduced \x2 greater than 1.4). Letting the metal abundance vary as a free
parameter in the fits (with the relative abundances fixed at the solar ratio)
increased  the number of fits with acceptable \x2,
but the fit is still unacceptable for the majority of the spectra. The values of
the abundances suggested by the fits were uniformly subsolar ($\sim 0.1$).

Since a single thermal component model is clearly not an adequate 
representation of the data, we introduced a
second thermal component with a different absorbing column
density [RS-(RS) model; here and in the following models the parentheses indicate the absorbed
component]. We first again assumed solar abundances, and we then let the abundance vary in
both components. In both cases, the overall N$_H$ was free to vary above the
Galactic value. 

The two-component, solar abundance, model provided  a statistically significant
improvement in the fit (above the 99.9\% confidence level based on an
F-test for two additional parameters; Bevington \& Robinson, 1992) in
all but 6  cases, over the single component, solar abundance model.
 In most cases the overall absorbing column is very close to the Galactic
\nh~ and the low temperature component has a
typical kT of 0.2-0.3~keV.  The second component is typically
seen through a larger column density (\nh$\sim2-8\times 10^{21}$~\cmsq), and
has a higher  temperature than the unabsorbed component (kT$\sim0.7-0.8$~keV).
 The results from these fits are presented in Table~1: Column (1) 
gives the data extraction region following the notation of fig.~3b, 
Column (2) gives the
net number of counts in each region (together with the statistical error), Column (3) gives
the  overall \nh~ in units of $10^{22}$~\cmsq, Column (4)
gives the temperature for the first component, while Columns (5)
and (6) give the column density and the temperature for the second
(absorbed) component. Column (7) gives the ratio of the emission
measures of the two best-fit thermal components. Column (8) gives the \x2 and the number of degrees of
freedom (dof). For the regions where a single component model gives an acceptable
fit only the parameters for this model are listed. 
Fig.~4 presents the 99\% confidence contours for the two
interesting parameters \nh\ and kT. Red and blue contours identify
the low and high temperature components respectively.
The crosses mark the best fit values for each component; if 
 the crosses are not shown the best fit values fall
outside the range of the plot.
Here, as in all the spectral fit tables, the confidence regions are
calculated usind the $\Delta$\x2 method, therefore they are a strong constraint
only in the case of acceptable fits (i.e. when the model is a reliable 
representation of the data). Regions 2C, 2E, 3E and 5C have unacceptable \x2 in
the solar abundance fit. These areas are part of the luminous emission
regions addressed in greater detail in \S3.2. 
In two instances in Table~1 (and Table~2) the best-fit
kT of the low temperature component is below 0.1~keV (i.e. outside the range of ACIS).
However, in these cases the upper bound is typically well within the instrumental range.
These results should be considered conservatively as upper limits on the temperature.

The results of the fits with the abundance as a free parameter in each thermal component
are listed in Table~2. Although the \x2\ are slightly improved, the fits are still not
acceptable in the bins with a larger number of counts. The best-fit temperatures, although
not identical, are in the general range of those obtained for the solar abundance fits.
The abundance is either ill-defined or subsolar [abundances are listed in
Cols. (4) and (7)]. 
Whenever a minimum is found, we list it with the errors. In some cases, a minimum is found,
but the confidence region extends outside one of the fit boundaries (0-5 for Z; 0-64~keV for 
kT; $> 3.4 \times 10^{20} cm^{-2}$, the line of sight value, for the total N$_H$). In these
cases, we list both the value of the parameter in question at the minimum \x2 
and the bound that can be determined from the explored grid. If no minimum is found, we 
give the bound calculated on the basis of the fitting grid boundary where the lowest 
value of \x2 is registered. In the case of the abundance, large values typically occur in
unconstrained cases, either because a minimum was not found in the \x2 distribution,
or, if a minimum was found, because the error is of the order of magnitude of the
best fit value. Some values of N$_H$ returned by the fit (e.g. for regions
5-D and 5-E) are extremely large, but also have comparably larger errors, indicating
that they are likely to be a consequence of poor statistics. In the case of region 5-E,
the extreme value of N$_H$ is not found when other models are adopted (e.g. see Table~1).
Similarly, the ratios of best-fit emission measures (both Table~1 and 2) show a few extreme values.
This occurs for regions 2-A,  5-E (both tables), and 2-E, 5-D (Table~2), 
indicating that one of the two components is likely to be an artifact of the fit.

 We also fitted the
data with a composite thermal plasma + power-law model (RS+PO). The rationale for this
model is that there could be a significant unresolved X-ray binary component to the 
emission, with fluxes below the detection threshold
of our data (see Paper II; the corresponding luminosity is
near $10^{38} \rm ergs~s^{-1}$). This model gave consistently  worse \x2 than 
the two-thermal-plasma model.

In summary, we conclude that this first fitting exercise strongly suggests a complex
temperature distribution in the ISM, at least in all the regions with higher
count statistics.

\subsection{Four Luminous Emission Regions}

Although the spectral grid gives a quantified version of the color image, showing
spectral variability throughout the soft emission, and suggesting a multi-temperature ISM,
the division of the data in a grid cuts some of the features of the ISM in a way that
it is not optimal. We have therefore selected four regions, identified in fig.~3b,
for further study. These regions are associated with the two
nuclei and the chain of hotspots to the  West of the northern nucleus (NGC~4038).
In fig.~3b they are marked as N for the northern nucleus
(NGC~4038), S for the southern nucleus (NGC~4039), and R1 and R2 for
the two hotspots to the West of the northern nucleus. The emission peak
in the S region was excised, because it corresponds to a point-like source
(Paper~II). 

We repeated the fitting procedures used for the binned array, using 
again on these regions the RS+(RS) and RS+(PO) model, with both solar abundance
and with the abundance as a free parameter. 
In the assumption of solar abundance, neither  model  gave an acceptable
fit to the spectrum of any region  but R2, which is also the region with the smallest 
number of spectral counts (see Table~3). We therefore did not pursue further analysis
for R2. The results from these fits are summarized in
Tables~3 and 4. 

Again, the introduction of the abundance as a free parameter leads to improved,
but still not good, fits. Preferred values of the abundance are subsolar.
This results is understandable if we look at fig.~5, 
that shows the best fit RS+(RS) solar abundance model 
 together with the residuals for these four regions. The model soft component
 tends to have a more pronounced peak in the region below 1~keV than the data.
 This is the regions of more prominent Fe-L emission.
With the exclusion of the RS+[PO] model for the two nuclei, the improvement
of the fit over the solar abundance models was statistically significant
above the 99\% confidence level based on an F-test.

Fits to 3-component models,
with fixed solar abundance (Table~5), give improved \x2 at a level similar
to those of the free abundance 2-component fits. These models
consist of two thermal RS components and either a power-law (PO) or a third RS component.
One of the RS components is either absorbed
by the same column as the power-law [RS+(RS+PO) model] or just by the
foreground  column [RS+RS+(PO) model]. Similarly, in the 3-RS component models,
 we also explored the possibility that the second
thermal component is associated either with the foreground or with the
obscured thermal component: these are noted as  RS+RS+(RS) and
RS+(RS+RS) respectively. 

All models  gave similar 
\x2, with the exception of the southern nucleus, where the 2-RS+PO models
are slightly favored. 
The best fit
temperatures for the two thermal components are kT$\sim0.2-0.3$~keV and
kT$\sim0.7-0.9$~keV respectively,
and the slope of the power-law is $\Gamma\sim3$. These power-laws are steeper
than expected from unresolved X-ray binaries (where $\Gamma\sim 2$
are more the norm; e.g. Schulz 1999, Yokogawa et al 2000).
The results from the three thermal
component models are also presented in Table~5.
While two of the components are in the range found by the previous fits, the third
component tends to be rather hard, and could be suggestive of unresolved hard sources.

The range of models used
are still not adequate to fit the spectra with the largest number of counts 
(southern nucleus and region R1, in that order).
We further explored the worst-fit, highest counts southern nucleus,
by allowing for the abundance to vary in the RS+(RS+PO) model.
However, this did not help the fit (\x2=95.8/63), and it led to totally
unconstrained results for the power-law and the abundance of the soft component.
We also divided the southern nucleus region in two sub-regions, one including
the E-W bar-like feature (see fig.~2), but we did not
get appreciably different results within statistics; the results are
consistent with those of Table~5. 

These result suggest complex emission regions. 
The different fits, both with solar abundance and
with the abundance as a free parameter, suggest two principal temperatures
of the hot ISM:
$\sim0.2-0.3$~keV and $\sim 0.6-0.8$~keV.
The abundance is either ill-defined or subsolar. Subsolar abundances were
also reported by Sansom et al (1996) from the spectral analysis of the integrated ASCA
spectrum of The Antennae. 
At this point, we believe that one should be cautious in over-interpreting
these possible subsolar abundances, given that the complexity of the emission regions, and the
relative low spectral resolution of the CCD data could conspire in giving erroneously
lower abundances (see e.g. demonstration in Kim \& Fabbiano 2003; Weaver et al 2000;
Matsushita et al 1997). Moreover, the elemental abundances were 
fixed at the solar ratio, and this may not hold in a starburst regions with a prevalently 
young and evolving massive stellar population, where Iron may be under-abundant, 
because of the prevalence of SN~II. Finally, uncertainties in the line emission
models themselves may conspire in giving an erroneously higher continuum, if the data are 
not of sufficient spectral resolution. Given the obvious spectral complexity 
of this ISM, and the relatively limited spatial resolution, resulting from the need
to integrate the spectra over sizeable areas, we believe that delving deeper in the
abundance problem is not warranted by these data.

In Table~6 we give the emission measures for the fits of Table~4 (EM\footnote{$\rm{EM
= \int{n^{2}dV}}$ where $\rm{n}$ is the density, V is the volume of
the gas}) and power-law (PO)
normalization for the three component models,
together with   
the absorption-corrected luminosity in the 0.1 -
10.0~keV band for each component.

\section{Low-Resolution 4cm and 6cm Radio Maps }

   Since we do not see sharp features in the hot ISM, 
   possibly because of the relatively low statistics of the X-ray data,
we have 
re-analyzed radio wavelength observations
to produce images of the diffuse radio emission, on 
scales similar  to those of our true-color X-ray map (fig.~2).

   VLA\footnote{The Very Large Array (VLA) is operated by the National
Radio Astronomy Observatory (NRAO), which is a facility of the National
Science Foundation operated under cooperative agreement by Associated
Universities, Inc.} 
archival data (20,  and 6cm) were used together with 6cm
observations from Neff \& Ulvestad (2000)  to
produce images sensitive to a range of spatial scales.  
The standard source 3C286 was used to set the flux density scale, 
and the source 1159-2148    was used as a phase calibrator.     
Each data set was edited, calibrated,
imaged, and then self-calibrated using standard image processing 
techniques and the Astronomical Image Processing System (AIPS).
Data sets observed using B1950 coordinates were precessed
to the J2000 system.  Data sets from different arrays were then 
combined and self-calibrated.  Data weighting was adjusted
to produce images of the desired angular resolution.
Image deconvolution (CLEANing) included the removal of four
background radio sources within the telescope's primary beam response. 
Because our final images are made from multiple data sets, some 
with unknown information about weather during the observations, we 
conservatively assume a flux calibration error of 5\%.  Absolute position 
errors of the brighter radio sources are less than 0.15\arcsec.  
20 and 6cm images are shown in fig.~6, with a restoring
beam size similar to that of the X-ray color map shown in fig.~2. 
       The resulting images were low-pass filtered to remove compact
radio sources (supernova remnants and HII regions), using a box
7.6 $\times$ 7.6 arcsec.  The smoothed images were combined to form 
a spectral index image (spectral index $\alpha$, where 
$S_{\nu} \propto \nu^{\alpha})$, 
using regions with S/N of 10 or better in the smoothed images (fig.~7).

\section{Discussion}

\subsection{Properties of the Hot ISM}

Fabbiano et al (2001) reported that
the {\it Chandra} image of \n4038\ shows extended, soft 
emission features with dimensions
ranging from a few arcseconds (few hundred parsecs) to the size of the
entire galaxian system or larger. In \S2 and \S3 we have taken a 
closer look at this emission and at its spectral properties. 
Here we discuss the implications of these results for the
properties of the hot ISM.

We used the results of the spectral analysis to
estimate the physical parameters of the
ISM in the two nuclear regions and in the
most intense well-defined emission region in the N-W arm 
(R1, see fig~3b and Table~5), and 
to infer the supernova rate in these regions. 
We used the emission
measures from the fit to the RS + (RS + PO) model (see Table~6).
In this model we assume
a cooler external unabsorbed ISM layer surrounding hotter ISM
and harder sources of emission embedded in a cooler/dusty absorbing
medium. This picture agrees with the true-color X-ray map, and 
suggests a hotter inner superbubble, heated by early-type stellar
winds and supernovae, with an outflowing and cooling upper layer of 
ISM. 

Our estimates are given in Table~7, for both low-kT (subscript 1), and
high-kT (subscript 2) thermal components (from $EM_1$ and $EM_2$ of Table~6).
For these calculations, we have assumed 
cylindrical emitting volumes of base given by the count extraction
areas of Table~2. Although it is likely that hot absorbed emission
originates from inner regions in a superbubble, with our 
data it is not possible to make an informed discrimination
between the different geometries (even in projection on the
sky) of the emitting regions. 
We assumed a height for the emitting cylinders  of either 200~pc or 1~kpc.
The thinner cylinder would correspond to an emission region of
the approximate depth of a spiral disk, while the thicker cylinder
assumes a scale-height more consistent with an outflowing hot ISM.
We give both estimates in Table~7. 
We estimated thermal energy 
and cooling times following Tucker (1975) and using the intrinsic
luminosities from Table~6. Thermal pressure in the assumption of 
electron and proton components in thermal equilibrium is $p = 2nkT$.

Supernova rates were derived
as in Heckman et al (1996; see also Fabbiano et al 1997 for a similar approach 
applied to the ROSAT HRI image of the Antennae),
which assumes that most of the mechanical energy supplied
to the ambient medium in the star-forming regions
is from supernovae. Following Heckman et al (1996), we adopt
the relation for a superbubble: $E_{th} = 1.4 \times 10^{57} L_{mech, 43} t_7
\rm ergs~s^{-1}$, whre $L_{mech, 43} $ is the mechanical energy supplied to the ambient
medium in the star forming region from supernovae, and $t_7$ is the
age of the starburst in units of 10$^7$~yr. $L_{mech } = R_{SN} \times E_{SN}$, 
where $R_{SN}$ is the supernova rate and
$E_{SN} = 10^{51}$~ergs is the energy released in a supernova explosion.
From this, we obtain an estimate of the supernova rate $R_{SN}$, equating
$E_{th}$ to our estimates of Table~7.

As shown by the X-ray/H$\alpha$ comparison in Paper~I,
in the nuclear regions the hot and warm ISM distributions
closely follow each other, suggesting that the two gases are
intermingled, as for example in the gaseous outflows of
the nearby starburst nucleus M82 (Watson et al 1984).
In this case the X-ray emitting volume may well be a fraction ($\eta$)
of the one adopted here ($\eta=1$ was assumed in Table.~7). This would result in shorter cooling
times ($\tau_c \sim \eta^{1/2}$) and similarly smaller thermal energies
and supernova rates, while the thermal pressure would increase as $p \sim \eta^{-1/2}$. 

The other region in Table~7 (R1) was
identified in Paper~I as a superbubble, with a hot core
surrounded by H$\alpha$ filaments. As suggested in Paper~I, R1 may be 
a more spectacular example of the superbubbles seen in more
nearby galaxies such as the LMC and M101 (e.g. Wang \& Helfand 1991; Williams \&
Chu 1995;
see reviews in Chu 2000 and MacLow 2000).

We derive cooling times in the $10^7-10^8$~yr range 
and total masses of hot ISM in the $10^5-10^6$~$M_{\odot}$ for each region.
For the two nuclear regions, where a comparison is possible,
these hot gas masses are of
order 1/1000 of the mass in cold molecular gas (Wilson et al 2000).
The thermal pressures of the hot gas are of the order of a 
few $\times 10^{-11}$~dyn~cm$^{-2}$. For the two nuclear regions, these
pressures are remarkably similar to those that can be derived for the cold 
molecular clouds from the CO measurements of Zhu, Seaquist \& Kuno (2003),
suggesting a rough pressure equilibrium of the different phases of the ISM.
Using the values of densities and temperatures from Zhu et al, we obtain cold gas
pressures of 4.2$\times 10^{-11}$~dyn~cm$^{-2}$ and 3.1$\times 10^{-11}$~dyn~cm$^{-2}$
for the northern and southern nuclear regions, respectively. In both nuclei,
for our assumption of gas volumes, the hotter X-ray components may have larger pressures, and
create bubbles or even escape from the enveloping molecular clouds. 
However, we must remember that these estimates (both CO and X-rays) are only average values,
depending on a number of assumptions, including the emitting volume and the small
scale properties of the clouds (density and temperature structures) that cannot
be measured at present.

In these regions our estimates of 
the supernova rates are within factors of a few of those
estimated by Neff \& Ulvestad (2000) from compact radio sources,
on the basis of radio continuum observations at 6cm.
We stress that the estimates in Table~7 are only indicative,
given the uncertainties in the geometry of the emitting regions,
and in the spectral models.
If the 3-temperature RS+(RS+RS) model is used (see Table~5), the total emission measure 
would be larger for the northern nucleus. But even so, the 
estimates of the ISM parameters would not change much:  this would result in a factor of 
$< 2$ increase in the mass of hot ISM, and correspondingly
shorter  cooling time.
The estimates for the southern nucleus would not change 
appreciably; in this case the fit returns two low-temperature
components with temperature and EM  similar to those from the RS+(RS+PO) model, and
a higher temperature component, that may be related to harder
emission from supernova remnants and X-ray binaries.
In the case of R1, the results of this fit
are probably unrealistic, requiring an extremely luminous,
very low energy component.

\subsection{Large-Scale  Morphology of the Multi-Wavelength Diffuse Emission}

Fig.~8 shows a  comparison of the  X-ray
image of the diffuse emission  (likely to be gas shock-heated by supernovae and stellar winds),
with the HST WFPC-2 archival H$\alpha$ 
image (Whitmore et al 1999), that traces photoionization of the ISM by luminous O stars, 
and with the low-pass filtered 6cm radio continuum 
VLA image of fig.~6, to which both the low-energy spectrum of the warm and hot ISM, 
as well as non-thermal contributions from embedded supernova remnants, contribute. 
Fig.~9 compares the
 diffuse X-ray emission of fig.2 (here plotted in red) with CO (Wilson et al 2000) and HI (Hibbard et al 2001)
images. 

These comparisons give a fairly complete picture of the large scale 
distribution of the different components of the ISM, and also of the effect of
the extinction on the emission.
 
The regions where young stars are being formed in dusty molecular clouds, as
traced by the CO, are
also prominent in the radio continuum, that is not affected by 
cold-ISM/dust absorption. The radio continuum
appears to follow the H$\alpha$ reasonably well, with the exception
of the peak at the NE of the southern nucleus (the dusty star-forming `impact
region', Mirabel et al 1998; Wilson et al 2000; Zhang et al 2001). 
 The radio spectral index map (fig.~7) indicates three regions where
  thermal emission ($\alpha > 0.0$)is present, probably from HII regions: the thickest
  part of the overlap region to the NE of NGC~4039 nucleus, another
  HII region complex further N and E of that, and a superbubble structure
  at the southern edge of the arc extending SW from the NGC~4038 nucleus.
  Filamentary structures in the non-thermal emission are evident running
  roughly EW across the northern half of the system, and may indicate
  magnetic field shears or discontinuities, given that the non-thermal 
  emission in galaxies is from synchrotron.  
The X-ray emission has a distinct minimum in the `impact region',
which is likely to be due to absorption of the soft photons
(the blue area in fig.~2). 
Mid-infrared ISO data have shown that this is the regions of most intense star formation,
including stars as massive as 60~$M_{\odot}$; the extinction in this
region reaches $A_V = 70$~mag (Mirabel et al 1998).
The blue area in fig.~2 can be seen trailing up to the northern nucleus (NGC~4038), following
the region rich in dust lanes (see HST image in Whitmore et al 1999).

All wavebands trace well 
the spiral arm that appears to emerge on the SW side of the northern nucleus (NGC~4038),
swings counter-clockwise  and is easily traceable for about 90 deg in radio and
X-ray. At that point, NW of the nucleus, the radio arm dims but 
continues its swing for another 90 degrees, as does the H$\alpha$ arm, and the HI emission.
The diffuse X-ray emission instead is faint in this region. This region does not present 
any peculiarity in the mix of ages of stellar clusters (Zhang et al 2001),
by comparison with other regions where diffuse X-ray emission
is present, that would suggest a significantly older stellar population
lacking the means of heating the ISM to X-ray temperatures. 
Hibbard et al (2001) point out that the HI kinematics in this 
area suggest infall of HI onto NGC~4038.
It may be possible that the infalling cold gas interact with and
displace the hot gas. A general pattern of avoidance between cold
and hot ISM can be seen for example in the merger remnant NGC~5128
(Karovska et al 2002). 
From the velocity maps and $N_H$ in Hibbard et al (2001), 
we estimate an infall velocity for the HI cloud
$v \sim 100$~km/s, and a mass density $\rho \sim 1.6 \times 10^{-24}$~g~cm$^{-3}$, corresponding
to 1 H atom per cubic centimeter. The ram pressure of this cloud would then be
$\rho~v^2 \sim 1.6 \times 10^{-10}~ \rm dyn~cm^{-3}$, larger than the typical 
thermal pressure of the hot ISM (see Table~7). Our lowest estimate of the
ram pressure is $\sim 3 \times 10^{-11} \rm dyn~cm^{-3}$, comparable to the pressure
estimates for the hot ISM. However, the pressures in Table~7 are derived for the 
most luminous X-ray regions, so are an overestimate of the pressure in the
region of the northern spiral arm. We conclude that displacement of the hot ISM
by infalling cold neutral hydrogen cloud is possible.

Interaction between cold and hot ISM is also suggested
by the relative distribution of CO and X-ray emission. In the impact region, 
X-ray colors and spectra may suggest extinction as the
cause of the low X-ray surface brightness, although, given the uncertainties in
the fits, a situation with an embedded obscured hard component, and a `surface'
low-emission measure soft component, is also possible. However, 
the extinction interpretation would be in agreement with emission originating from
a region of star formation deeply 
embedded in the molecular clouds, since the CO emission peaks in this area.
Other CO/X-ray overlap or avoidance areas 
are not immediately ascribable to extinction. These regions include the X-ray 
gap in the NW spiral arm, and regions where the X-ray emission is instead 
intense: the `superbubble' in the NW arm discussed in Paper~I (regions R1 and R2
of fig. 3b), and the two 
nuclear regions. As we have shown in \S5.1, the pressures in the cold (CO) and hot
(X-ray) ISM are similar and suggest a situation where cold and hot ISM may be
commingled and in equilibrium.
Near the most intense region of diffuse X-ray emission at the southern nucleus (NGC~4039)
there is a marked minimum of both HI and CO. This may suggest a hot nuclear outflow
pushing aside the cold ISM, reminiscent of the morphology of
M82 and NGC~253 (Watson et al 1984, Fabbiano \& Trinchieri 1984).

\section{Summary and Conclusions }

This paper presents a first look at the complex hot ISM
of the Antennae galaxies, based on the first {\it Chandra}
observation of this merging system.

After producing an X-ray-color image,
that gives us an immediate visual representation of
the spectral complexity of the diffuse emission, we have
obtained a quantitative estimate of this complexity by means of
spectral fits. At least two temperatures (with different $N_H$)
are needed to fit the most luminous regions, including the two nuclei, 
and what appear to be luminous superbubbles (Paper~I) in the N-W star-forming arm of NGC~4038.
The two temperatures are in the vicinities of 0.3 and 0.7~keV, with 
the hotter component being more highly absorbed. However, the fits
(even adding a third component) are still not a good representation
of the data. We have explored the effect of non-solar abundances in these
fits. While the fits improve for very low abundance values (as already
reported with ASCA, Sansom et al 1996), they are still not good.
Moreover, given the clear complexity of the emission field, suggested both
by the X-ray-color image and by the spectral results, spuriously low
value of the abundance may result because of a simplistic choice of
fit models (e.g. Weaver et al 2000; Kim \& Fabbiano 2003).

From these results we derive estimates of the hot ISM 
parameters for these three regions. 
We derive cooling times of $\sim 10^7-10^8$~yr 
and masses of hot ISM of $\sim 10^5-10^6$~$M_{\odot}$ for each region.
In the two nuclei, these masses are much smaller than those present in
molecular gas (Wilson et al 2000), but the thermal pressures of the
two phases of the ISM are comparable, suggesting equilibrium.
However, this conclusion depends on crucial assumption on the spatial 
distribution and filling factor of the ISM. It is also possible that, at
least for the hottest component of the hot ISM, the thermal pressure
may exceed that of the cold medium.
For the two nuclear regions, where a comparison is possible,
the supernova rates are of the order of magnitude of those 
estimated by Neff \& Ulvestad (2000) for the same regions
on the basis of radio continuum observations at 6cm.

We have compared the distribution of the diffuse X-ray emission
with that of the 6cm radio continuum,
H$\alpha$ (using an archival HST WFPC-2 image), HI (Hibbard et al 2001)
and CO (Wilson et al 2000). While there is a general resemblance between
different wavebands, that may be related to a common link with
star formation phenomena (Zhang et al 2001), this comparison shows significant
differences. In particular, it shows very clearly the
effect of obscuration on the X-ray emission, that has a minimum
(where only hard photons escape) coincident with the large
radio continuum and CO peaks in the impact region of the two galaxies.
We also note a displacement of the diffuse X-ray emission on the northen side
of the system, relative to the spiral arm visible in H$\alpha$, radio continuum
and HI. The HI velocities suggest infall in this region (Hibbard et al 2001),
that results in a ram pressure exceeding the thermal pressure of the hot gas.
This could explain the displacement of the X-ray emission towards the south.

These results demonstrate the richness of the ISM
of the Antennae galaxies and justify further in dept-studies. 
A much deeper X-ray image is being 
collected with {\it Chandra} that will allow in time far more
detailed studies of individual regions.

\acknowledgments

We thank the CXC DS and SDS teams for their efforts in reducing the data and 
developing the software used for the reduction (SDP) and analysis (CIAO).
We thank our colleagues J. Hibbard and C. Wilson for providing
the digital versions of their data that were used in this work,
and D. Burke for help in producing fig.~6.
The archival {\it Chandra} ACIS-S data used for this work 
was originally obtained as part of the GTO program (PI S. S. Murray).
This work was supported by NASA contract NAS~8--39073 (CXC).


\begin{figure}
\caption{Multi-color image of the Antennae prior to point source
removal.  Point sources to be removed are indicated by white
ellipses. The color scale has been modified to enhance faint features.}\label{fg:ptsrc_img}
\end{figure}

\begin{figure}
\caption{Multi-color image after point source removal.  The color
scale has been modified to enhance faint features.}\label{fg:diff_img}
\end{figure}

\begin{figure}                                          
\caption{ (a) Raw full band image of the Antennae after extracting
the point sources indicated by their ellipses and numbered following
the notation of Paper~II. (b) Adaptively smoothed (0.3-6.0)~keV band image 
of the diffuse emission of the Antennae, showing the regions used to extract 
the X-ray spectra.    }
\end{figure}

\begin{figure}                                          
\caption{ Confidence contours for the RS+(RS) fits for the regions
shown if Fig.~1b. Red contours corresponbd to the unabsorbed temperature
component while blue contours correspond to the absorbed
component. The crosses mark the best fit parameters.}
\end{figure}

\begin{figure}
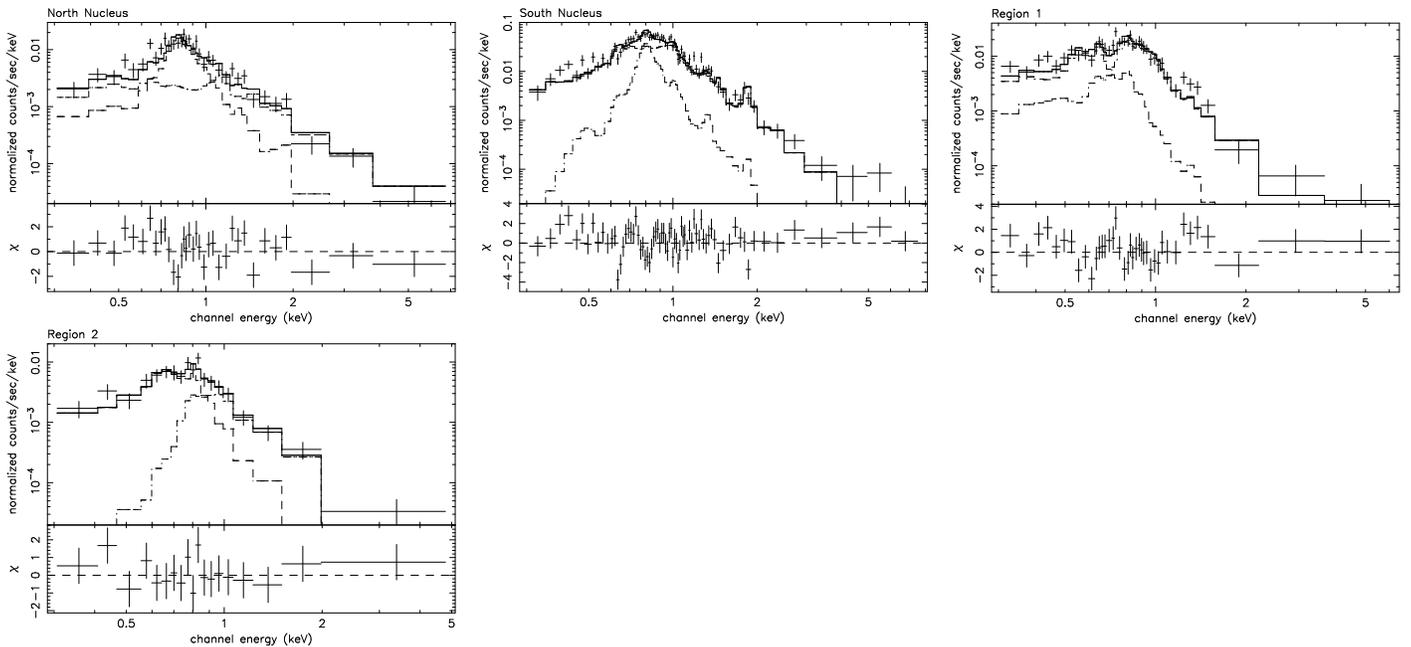
                                          
\rotatebox{270}{\includegraphics[height=6.0cm]{f5a.eps}} 
\rotatebox{270}{\includegraphics[height=6.0cm]{f5b.eps}}
\rotatebox{270}{\includegraphics[height=6.0cm]{f5c.eps}} 
\rotatebox{270}{\includegraphics[height=6.0cm]{f5d.eps}}
\caption{Spectra of the Northern and Southern nucleus and Regions 1 and
2 (see text for details) together with the best fit solar abundance RS+(RS) model and
the residuals in $\sigma$. The dashed and dot-dashed lines show the
two thermal components.  }
\end{figure}   

\
\clearpage
\begin{figure}
\caption{Image of the low-filtered diffuse radio emission in 
NGC4038/4039 at 6cm(left) and at 20cm(right).  
Both images have been restored with a beam 6.59\arcsec $\times$ 4.25\arcsec, 
and smoothed with a 7.6\arcsec $\times$ 7.6\arcsec low-pass filter 
to remove the compact radio sources. 
The 20cm image has an rms of 52mJy/beam, the 6cm image has an 
rms of 18mJy/beam.  
} 
\end{figure}

\begin{figure}
\caption{Gray-scale representation of the diffuse emission's 20-6cm spectral index
distribution in NGC4038/9.  This image shows only regions where
both input images have a flux density of at least 10$\sigma$.  
The majority of the diffuse emission has a spectral index $\alpha < $-0.5
(S$_{\nu}\propto\nu^{\alpha}$) implying non-thermal synchrotron emission from 
relativistic electrons.  Three regions have $\alpha > -0.5$; all are
locations of significant current star-formation.   Filamentary structures
in the non-thermal emission probably indicates shocks and/or regions
of magnetic field compression.  
} 
\end{figure}

\begin{figure}
\caption{Radio/X-ray  image of the diffuse emission in the Antennae. The radio 
image (green) is from the 6~cm  low-pass filtered image in fig.~6.
The X-ray image (red) is from the (0.3-6.0)~keV 
band and is adaptively smoothed. The countours give sketchy outlines of the 
H$\alpha$ emission from the HST WFPC2 data (see Paper~I), and 
identify the most active HII regions.
The circles in the bottom right of the image 
indicate the resolution of the radio and X-ray bands. In the case of the X-ray band they 
indicate the minimum and the maximum smoothing scales.
}
\end{figure}

\begin{figure}
\caption{ A combined  HI (red), CO (green), and X-ray (blue)  image of the 
Antennae. The HI data are from the maps of Hibbard \etal (2001) and the CO 
data are from Wilson \etal (2001). The X-ray data are in the (0.3-6.0)~keV 
band.  The circles in the bottom right of 
the figure indicate the resolution of each image. For the X-ray band 
they indicate the minimum and the maximum smoothing scales. }
\end{figure}

\clearpage

\begin{deluxetable}{lcccccccccccccccc}
\tabletypesize{\scriptsize}
\tablecolumns{17}
\tablewidth{0pt}
\tablecaption{Spectral Fits for the Temperature Map ($\rm{Z=Z_{\odot}}$)}
\tablehead{\colhead{Region}& \colhead{Net counts}& \colhead{$\rm{N_{H,tot}^{a}}$}&
\colhead{kT$^{b}_{1}$}& \colhead{$\rm{N_{H}^{c}}$}& \colhead{kT$^{d}_{2}$}&
\colhead{$\rm{EM_{1}/EM_{2}}$}& \x2 (dof)\\
\colhead{ (1) } &\colhead{ (2)}
&\colhead{(3)} &\colhead{(4)} &\colhead{(5)} & \colhead{ (6) }  &\colhead{ (7)}&\colhead{ (8)}}
\startdata
1-A$^{e}$ & $   33.4  \pm  13.1  $ &$  0.22^{+ 0.44 }_{- 0.19 } $ & $
0.1 (<0.2) $ &                             &                             &       & 5.4 (8)  \\[0.20cm]
1-B$^{e}$ & $  135.1  \pm  16.7  $ &$  0.60^{+ 0.17 }_{- 0.27 } $ & $ 0.14^{+ 0.06}_{- 0.07} $ &                             &                             &       & 13.3 (14)  \\[0.20cm]
1-C & $  285.4  \pm  20.6  $ &$ 0.034 (<0.1)              $ & $ 0.28^{+ 0.03}_{- 0.04} $ &  $   < 0.17               $ &  $ 5.74^{+ 0.26}_{- 5.67} $ &  0.52 & 22.6 (19)  \\[0.20cm]
1-D & $  247.4  \pm  19.4  $ &$ 0.034 (<0.09)             $ & $ 0.20^{+ 0.04}_{- 0.04} $ &  $ 0.30^{+ 0.28}_{- 0.30} $ &  $ 0.70^{+ 0.10}_{- 0.13} $ &  0.79 & 9.5 (17)  \\[0.20cm]
1-E$^{e}$ & $  130.2  \pm  16.6  $ &$  0.71^{+ 0.86 }_{- 0.19 } $ & $ 0.14^{+ 0.03}_{- 0.08} $ &                             &                             &       & 15.0 (14)  \\[0.20cm]
2-A & $  207.9  \pm  18.5  $ &$ 0.94^{+ 0.33}_{- 0.26}    $ & $ 0.09^{+ 0.03}_{- 0.02} $ &  $                  >3.19 $ &  $ 6.6   (>0.7)           $ &  12300& 11.1 (15)  \\[0.20cm]
2-B & $  767.0  \pm  30.1  $ &$ 0.034 (<0.063)            $ & $ 0.26^{+ 0.03}_{- 0.04} $ &  $ 0.38^{+ 0.15}_{- 0.15} $ &  $ 0.69^{+ 0.06}_{- 0.07} $ &  0.37 & 44.3 (38)  \\[0.20cm]
2-C & $ 1520.7  \pm  40.7  $ &$ 0.018^{+ 0.027}_{- 0.016} $ & $ 0.31^{+ 0.20}_{- 0.02} $ &  $ 0.67^{+ 0.13}_{- 0.11} $ &  $ 0.71^{+ 0.50}_{- 0.03} $ &  0.25 & 123.0 (58)  \\[0.20cm]
2-D & $  554.4  \pm  26.4  $ &$ 0.034 (<0.087)            $ & $ 0.26^{+ 0.02}_{- 0.03} $ &  $ 0.38^{+ 0.28}_{- 0.35} $ &  $ 0.85^{+ 0.13}_{- 0.13} $ &  0.60 & 44.1 (31)  \\[0.20cm]
2-E & $  549.7  \pm  26.4  $ &$ 0.034 (<0.056)            $ & $ 0.26^{+ 0.04}_{- 0.03} $ &  $ 0.37^{+ 0.22}_{- 0.37} $ &  $ 0.72^{+ 0.28}_{- 0.07} $ &  0.89 & 48.7 (29)  \\[0.20cm]
3-A & $  928.2  \pm  32.5  $ &$ 0.034 (<0.054)            $ & $ 0.25^{+ 0.05}_{- 0.03} $ &  $ 0.41^{+ 0.12}_{- 0.17} $ &  $ 0.70^{+ 0.07}_{- 0.06} $ &  0.39 & 44.7 (46)  \\[0.20cm]
3-B & $  866.5  \pm  31.4  $ &$  0.04^{+ 0.05 }_{- 0.01}  $ & $ 0.25^{+ 0.03}_{- 0.03} $ &  $ 0.70^{+ 0.12}_{- 0.14} $ &  $ 0.72^{+ 0.06}_{- 0.04} $ &  0.15 & 54.7 (42)  \\[0.20cm]
3-C & $  751.0  \pm  29.9  $ &$  0.79^{+ 0.26 }_{- 0.19 } $ & $ 0.11^{+ 0.03}_{- 0.02} $ &  $                 >0.12  $ &  $ 0.61^{+ 0.08}_{- 0.19} $ &  209  & 39.9 (41)  \\[0.20cm]
3-D & $  694.6  \pm  28.6  $ &$ 0.034 (<0.049)            $ & $ 0.25^{+ 0.03}_{- 0.03} $ &  $ 0.44^{+ 0.15}_{- 0.18} $ &  $ 0.69^{+ 0.08}_{- 0.06} $ &  0.35 & 40.6 (36)  \\[0.20cm]
3-E & $  794.7  \pm  30.5  $ &$ 0.034 (<0.052)            $ & $ 0.21^{+ 0.03}_{- 0.02} $ &  $ 0.24^{+ 0.15}_{- 0.24} $ &  $ 0.69^{+ 0.12}_{- 0.06} $ &  0.77 & 58.9 (42)  \\[0.20cm]
4-A$^{e}$ & $  202.0  \pm  18.3  $ &$  0.81^{+ 0.11 }_{- 0.22 } $ & $ 0.15^{+ 0.14}_{- 0.22} $ &  $                        $ &  $                        $ &       & 15.1 (17)  \\[0.20cm]
4-B & $  368.0  \pm  22.1  $ &$ 0.034 (<0.11)             $ & $ 0.30^{+ 0.05}_{- 0.06} $ &  $ 0.80^{+ 0.20}_{- 0.35} $ &  $ 0.88^{+ 0.12}_{- 0.12} $ &  0.09 & 23.2 (23)  \\[0.20cm]
4-C & $  483.1  \pm  24.9  $ &$  0.06^{+ 0.08}_{- 0.02  } $ & $ 0.26^{+ 0.04}_{- 0.04} $ &  $ 0.26^{+ 0.28}_{- 0.26} $ &  $ 0.79^{+ 0.09}_{- 0.10} $ &  0.58 & 16.6 (26)  \\[0.20cm]
4-D & $  264.6  \pm  20.3  $ &$ 0.034 (<0.94)             $ & $ 0.26^{+ 0.05}_{- 0.04} $ &  $ 0.41^{+ 0.42}_{- 0.41} $ &  $ 0.78^{+ 0.22}_{- 0.16} $ &  0.60 & 12.6 (18)  \\[0.20cm]
4-E & $  127.2  \pm  16.5  $ &$ 0.034 (<0.1)              $ & $ 0.28^{+ 0.03}_{- 0.03} $ &  $ 0.99^{+ 47.0}_{- 1.00} $ &  $ >2.24                  $ &  0.33 & 11.3 (11)  \\[0.20cm]
5-A & $  141.7  \pm  17.0  $ &$   1.4^{+ 1.1  }_{- 0.54 } $ & $ 0.07^{+ 0.05}_{- 0.03} $ &  $ 8.45^{+ 4.45}_{- 8.45} $ &  $ 0.35^{+ 20.6}_{- 0.27} $ &  654.4& 8.3 (11)  \\[0.20cm]
5-B & $  714.4  \pm  29.1  $ &$ 0.034 (<0.079)            $ & $ 0.30^{+ 0.03}_{- 0.06} $ &  $ 0.37^{+ 0.13}_{- 0.18} $ &  $ 0.83^{+ 0.09}_{- 0.09} $ &  0.28 & 46.5 (38)  \\[0.20cm]
5-C & $ 2101.0  \pm  47.3  $ &$ 0.034 (<0.047)            $ & $ 0.30^{+ 0.02}_{- 0.02} $ &  $ 0.42^{+ 0.10}_{- 0.10} $ &  $ 0.78^{+ 0.05}_{- 0.05} $ &  0.32 & 104.5 (68)  \\[0.20cm]
5-D & $  374.2  \pm  22.7  $ &$ 0.034 (<0.06)             $ & $ 0.30^{+ 0.05}_{- 0.12} $ &  $ < 0.29                 $ &  $ 0.71^{+ 0.16}_{- 0.13} $ &  1.76 & 28.0 (22)  \\[0.20cm]
5-E & $  131.9  \pm  16.5  $ &$  0.39^{+ 0.36 }_{- 0.36 } $ & $ 0.28^{+ 0.12}_{- 0.10} $ &  $ >17.9                  $ &  $ 0.22^{+ 0.01}_{- 0.07} $ &  $<$0.0001 & 11.2 (11)  \\[0.20cm]
6-A$^{e}$ & $  116.2  \pm  16.1  $ &$ 0.034 (<0.126)            $ & $ 0.27^{+ 0.03}_{- 0.05} $ &  $                        $ &  $                        $ &       & 11.7 (13)  \\[0.20cm]
6-B$^{e}$ & $  150.6  \pm  17.1  $ &$  0.89^{+ 0.22 }_{- 0.18 } $ & $ 0.12^{+ 0.03}_{- 0.03} $ &  $                        $ &  $                        $ &       & 14.1 (15)  \\[0.20cm]
6-C & $  293.9  \pm  20.7  $ &$  0.23^{+ 0.23 }_{- 0.20 } $ & $ 0.13^{+ 0.06}_{- 0.06} $ &  $ < 0.29                 $ &  $ 0.48^{+ 0.15}_{- 0.17} $ &  3.90 & 15.5 (18)  \\[0.20cm]
6-D & $  139.7  \pm  16.8  $ &$ 0.034  (<0.15)            $ & $ 0.33^{+ 0.11}_{- 0.07} $ &  $      <0.52             $ &  $ 1.26^{+ 0.77}_{- 0.26} $ &  0.52 & 8.2 (11)  \\[0.20cm]
\enddata
\tablenotetext{a,c}{ Absorbing column density for total spectrum (a)
and the second thermal component (c) in units of
$10^{22}~\rm{cm^{-2}}$. Quoted errors are at the 90\% confidence level
for one interesting parameter.}
\tablenotetext{b,d}{ Temperature for the unabsorbed (b) and the
absorbed (d) thermal component in keV. Quoted errors are at the 90\% confidence level
for one interesting parameter.}
\tablenotetext{e}{ Fits with only one thermal component.}
\end{deluxetable}

\begin{deluxetable}{lcccccccccccccccc}
\tabletypesize{\scriptsize}
\tablecolumns{17}
\tablewidth{0pt}
\tablecaption{Spectral Fits for the Temperature Map (free Abundance)}
\tablehead{\colhead{Region}& \colhead{$\rm{N_{H,tot}^{a}}$}&
\colhead{kT$^{b}_{1}$}&\colhead{$\rm{Z^{e}}$}&
\colhead{$\rm{N_{H}^{c}}$}& \colhead{kT$^{d}_{2}$}&
\colhead{$\rm{Z^{f}}$}& \colhead{$EM_{1}/EM_{2}$} & \colhead{\x2 (dof)}\\
\colhead{ (1) } &\colhead{ (2)}
&\colhead{(3)} &\colhead{(4)} &\colhead{(5)} & \colhead{ (6) }  &\colhead{ (7)}&\colhead{ (8)}&\colhead{ (9)}}
\startdata
1-A$^{g}$ & $0.58^{+1.01}_{-0.54}$ & $0.05^{+0.18}_{-0.03}$ & $0.02(<5.0)$           &                         &                        &                         &           &3.3/7\\[0.20cm]   
1-B$^{g}$ & $0.24 (<0.678)$	   & $0.22_{0.102}^{+0.10}$ & $0.03^{+0.26}_{-0.021}$&                         & 			&                         &           &9.7/13\\[0.20cm] 
1-C       & $0.15_{-0.12}^{+0.21}$ & $0.28_{-0.08}^{+0.05}$ & $0.04_{-0.03}^{+4.96}$ &  $<9.02$        	       &$64.0^{h}$              & $5.00^{h}$              &    19.02  &20.0/17 \\[0.20cm]      
1-D       & $0.03_{-0.00}^{+0.08}$ & $0.21_{-0.05}^{+0.05}$ & $0.24_{-0.19}^{+4.76}$ &  $0.36_{-0.36}^{+0.30}$ &$0.70_{-0.13}^{+0.15}$  & $4.00 (>0.26)$ 	  &    6.96   & 9.1/15 \\[0.20cm]         
1-E$^{g}$ & $0.19 (<1.05)$ 	   & $0.03 (< 0.24)       $ & $0.28^{+0.23}_{-0.19}$ &         &                        &                         &           &11.2/12\\[0.20cm]
2-A       & $0.74_{-0.45}^{+0.39}$ & $0.11_{-0.04}^{+0.13}$ & $0.03_{-0.03}^{+0.42}$ &  $0.85_{-0.85}^{+4.76}$ &$4.48 (>0.45)$ 		& $5.00^{h}$		  &    9012 & 7.8/13 \\[0.20cm]       
2-B       & $0.03_{-0.00}^{+0.06}$ & $0.27_{-0.05}^{+0.06}$ & $0.27_{-0.18}^{+4.73}$ &  $0.30_{-0.26}^{+0.28}$ &$0.70_{-0.07}^{+0.07}$  & $0.45 (>0.14)$ 	  &    0.70   & 42.5/36 \\[0.20cm]      
2-C       & $0.03_{-0.00}^{+0.05}$ & $0.31_{-0.03}^{+0.03}$ & $0.33_{-0.22}^{+4.67}$ &  $0.25_{-0.14}^{+0.31}$ &$0.71_{-0.03}^{+0.07}$  & $0.09_{-0.04}^{+0.24}$  &    0.24   & 111.5/56 \\[0.20cm]      
2-D       & $0.06_{-0.02}^{+0.06}$ & $0.27_{-0.05}^{+0.03}$ & $4.90_{-4.52}^{+0.10}$ &  $0.02_{-0.02}^{+0.36}$ &$0.89_{-0.12}^{+0.10}$  & $0.14_{-0.09}^{+0.20}$  &    0.05   & 38.9/29 \\[0.20cm]       
2-E       & $0.03_{-0.00}^{+0.05}$ & $0.30_{-0.03}^{+0.04}$ & $0.21_{-0.13}^{+0.41}$ &  $1.90_{-0.90}^{+1.02}$ &$0.09_{-0.01}^{+0.04}$  & $4.55_{-4.45}^{+0.45}$  &    0.0001   & 37.3/27 \\[0.20cm]       
3-A       & $0.06_{-0.02}^{+0.06}$ & $0.27_{-0.04}^{+0.06}$ & $0.13_{-0.07}^{+0.25}$ &  $0.40_{-0.34}^{+0.22}$ &$0.69_{-0.06}^{+0.10}$  & $1.29_{-1.10}^{+3.71}$  &    3.54   & 38.3/44 \\[0.20cm]       
3-B       & $0.13_{-0.09}^{+0.15}$ & $0.25_{-0.06}^{+0.06}$ & $0.07_{-0.05}^{+0.41}$ &  $0.64_{-0.27}^{+0.16}$ &$0.72_{-0.05}^{+0.07}$  & $1.12_{-0.86}^{+3.88}$  &    3.00   & 51.1/40 \\[0.20cm]       
3-C       & $0.29_{-0.16}^{+0.30}$ & $0.22_{-0.07}^{+0.07}$ & $0.02_{-0.02}^{+4.98}$ &  $<0.36$                &$0.77_{-0.10}^{+0.11}$  & $0.16_{-0.10}^{+1.38}$  &    14.68  &25.8/39 \\[0.20cm]        
3-D       & $0.03_{-0.00}^{+0.04}$ & $0.30_{-0.06}^{+0.06}$ & $0.11_{-0.05}^{+0.10}$ &  $0.53_{-0.32}^{+0.19}$ &$0.70_{-0.06}^{+0.09}$  & $4.99 (>0.39)$ 	  &    10.52  & 28.8/34 \\[0.20cm]       
3-E       & $0.03_{-0.00}^{+0.06}$ & $0.23_{-0.05}^{+0.04}$ & $1.38_{-1.28}^{+3.62}$ &  $<0.35$                &$0.71_{-0.08}^{+0.09}$  & $0.13_{-0.05}^{+1.49}$  &    0.11   & 45.1/40 \\[0.20cm]       
4-A$^{g}$ & $0.37_{-0.19}^{+1.13}$ & $0.27_{-0.20}^{+0.11}$ & $0.05^{+0.11}_{-0.04}$ &                         &			&                         &           &11.6/16\\[0.20cm] 
4-B       & $0.06_{-0.03}^{+0.20}$ & $0.35_{-0.09}^{+0.24}$ & $0.09_{-0.07}^{+0.69}$ &  $0.67_{-0.55}^{+0.37}$ &$0.91_{-0.14}^{+0.21}$  & $0.45_{-0.36}^{+4.55}$  &    0.52   & 19.7/21 \\[0.20cm]        
4-C       & $0.16_{-0.13}^{+0.17}$ & $0.26_{-0.07}^{+0.08}$ & $0.07_{-0.05}^{+4.93}$ &  $0.20_{-0.20}^{+0.37}$ &$0.79_{-0.11}^{+0.14}$  & $1.42_{-1.31}^{+3.58}$  &    16.47  & 14.0/24\\[0.20cm]         
4-D       & $0.03_{-0.00}^{+0.10}$ & $0.29_{-0.07}^{+0.07}$ & $0.20_{-0.15}^{+4.80}$ &  $0.52_{-0.52}^{+0.52}$ &$0.79_{-0.15}^{+0.24}$  & $4.99_{-4.74}^{+0.01}$  &    10.34  & 11.1/16 \\[0.20cm]        
4-E       & $0.03_{-0.00}^{+0.16}$ & $0.30_{-0.05}^{+0.10}$ & $0.13_{-0.10}^{+4.87}$ &  $5.93_{-5.93}^{+44.6}$ &$1.44 (>0.64)$ 		& $5.00^{h}$ 		  &    1.09   & 10.0/9 \\[0.20cm]         
5-A       & $1.12_{-1.02}^{+1.11}$ & $0.08_{-0.04}^{+0.32}$ & $0.03_{-0.03}^{+4.97}$ &  $10.9_{-10.9}^{+33.8}$ &$0.27_{-0.20}^{+2.57}$  & $0.03_{-0.03}^{+4.97}$  &    5.10   & 6.2/9 \\[0.20cm]          
5-B       & $0.09_{-0.06}^{+0.14}$ & $0.32_{-0.07}^{+0.08}$ & $0.13_{-0.09}^{+1.18}$ &  $0.40_{-0.40}^{+0.21}$ &$0.83_{-0.09}^{+0.14}$  & $3.36_{-3.07}^{+1.64}$  &    7.21   & 44.1/36 \\[0.20cm]        
5-C       & $0.03_{-0.00}^{+0.06}$ & $0.34_{-0.04}^{+0.04}$ & $0.41_{-0.30}^{+4.59}$ &  $0.13_{-0.13}^{+0.14}$ &$0.82_{-0.06}^{+0.06}$  & $0.18_{-0.06}^{+0.19}$  &    0.25   & 84.8/66 \\[0.20cm]        
5-D       & $0.05_{-0.01}^{+0.08}$ & $0.42_{-0.06}^{+0.10}$ & $0.13_{-0.05}^{+0.11}$ &  $89.1_{-83.6}^{+42.9}$ &$0.17_{-0.04}^{+0.04}$  & $4.83^{h}$ 		  &    $<$0.0001   & 21.9/20 \\[0.20cm]        
5-E       & $0.39_{-0.33}^{+0.33}$ & $0.28_{-0.10}^{+0.14}$ & $0.16_{-0.12}^{+4.85}$ &  $118._{-45.7}^{+27.0}$ &$0.17_{-0.03}^{+0.08}$  & $4.32^{h}$ 		  &    $<$0.0001   & 10.5/9 \\[0.20cm]           
6-A$^{g}$ & $0.034 (<0.16)$        & $0.31_{-0.07}^{+0.10}$ & $0.06_{-0.04}^{+0.18}$ &                         &                        &                         &           &6.4/12\\[0.20cm]
6-B$^{g}$ & $0.79_{-0.44}^{+0.38}$ & $0.12_{-0.03}^{+0.10}$ & $0.04^{h}$ 	     &                         &                        &                         &           &6.4/12\\[0.20cm]
6-C       & $0.14_{-0.09}^{+0.11}$ & $0.41_{-0.08}^{+0.12}$ & $0.10_{-0.04}^{+0.09}$ &  $10.8_{-10.8}^{+9.20}$ &$63.9^{h}$ 		& $0.88^{h}$ 	 	  &    5.30   & 18.1/16 \\[0.20cm]
6-D       & $0.03 (0.36)$	   & $0.36^{+0.27}_{-0.13}$ & $0.12^{h}$ 	     &  $<1.18$                &$0.31_{-0.75}^{+5.98}$  & $5.00^{h}$		  &    15.79  & 7.2/9 \\[0.20cm] 
\enddata
\tablenotetext{a,c}{ Absorbing column density for total spectrum (a)
and the second thermal component (c) in units of
$10^{22}~\rm{cm^{-2}}$. Quoted errors are at the 90\% confidence level
for one interesting parameter.}
\tablenotetext{b,d}{ Temperature for the unabsorbed (b) and the
absorbed (d) thermal component in keV. Quoted errors are at the 90\% confidence level
for one interesting parameter.}
\tablenotetext{e,f}{ Abundance for the unabsorbed (b) and the
absorbed (d) thermal component in keV. Quoted errors are at the 90\% confidence level
for one interesting parameter.}	
\tablenotetext{g}{ Fits with only one thermal component.}
\tablenotetext{h}{ Unconstrained parameter (1$\sigma$ range greater
than allowed range for the parameter).}
\end{deluxetable}

\begin{deluxetable}{lccccccccccc}
\tablenum{3}
\tablecolumns{11}
\tablewidth{0pt}
\tablecaption{ Spectral Fits for Selected Regions (Z=Z$\odot$) }
\tablehead{ \colhead{Region}& \colhead{RA (J2000)} &
\multicolumn{2}{c}{RS+(PO)} &  \multicolumn{2}{c}{RS+(RS)}   \\
\colhead{Net counts $\pm$ error }  &  \colhead{Dec (J2000)}    & \colhead{$\rm{N_{H,tot}}^{a}$ }  &\colhead{kT$^{b}_{1}$ } 
& \colhead{$\rm{N_{H,tot}}^{a}$ } & \colhead{kT$^{b}_{1}$ }  \\
\colhead{ }  &\colhead{Area (sq. arcsec) }  &\colhead{$\rm{N_{H}}^{c}$ }  &\colhead{ $\Gamma$} &
\colhead{$\rm{N_{H}}^{c}$ } & \colhead{kT$^{d}_{2}$} &  \\
\colhead{ }      &\colhead{ }      & \colhead{ } &\colhead{ \x2 (dof)}  &
 \colhead{ }  &\colhead{ \x2 (dof)}  &   \\
\colhead{ (1) } &\colhead{ (2)}
&\colhead{(3)} &\colhead{(4)} &\colhead{(5)} &\colhead{(6)}} 
\startdata
North Nucleus & 12: 01: 52.8 & $0.034 (<0.2)$ & $0.65^{+0.06}_{-0.08}$ & $0.034 (<0.045)$ & $0.65^{+0.03}_{-0.06}$ \\[0.2cm]
$604.7\pm25.4$& -18: 52: 08.5 &  $0.15 (<0.28)$ & $2.9^{+0.7}_{-0.5}$ & $ <0.015 $ &$3.04^{+1.4}_{-0.79}$  \\[0.2cm]
& 222  & & 38.9 (29) & & 52.3 (29)  \\[0.2cm]
South Nucleus & 12: 01: 53.9& $0.034 (<0.058)$ & $0.76^{+0.03}_{-0.03}$ & $0.034 (<0.045)$ &$0.32^{+0.02}_{-0.03}$  \\[0.2cm]
$2180.4\pm47.5$&  -18: 53: 08.5& $0.22^{+0.03}_{-0.09}$ & $3.4^{+0.7}_{-0.5}$ &$0.41^{+0.1}_{-0.09}$ &$0.79^{+0.05}_{-0.04}$  \\[0.2cm]
&  488  & & 138.7 (67) & & 125.1 (67)  \\[0.2cm]
Region~1 & 12: 01: 50.4 & $0.034 (<0.08)$ &$0.65^{+0.05}_{-0.05}$ & $0.034 (<0.05)$ & $0.20^{+0.05}_{-0.02}$ \\[0.2cm]
$780.3\pm28.9$& -18: 52: 19.1&  $0.12^{+0.09}_{-0.09}$ & $3.98^{+0.73}_{-0.64}$ & $<0.18$ & $0.78^{+0.06}_{-0.06}$   \\[0.2cm]
& 235 &  & 64.9 (34) & & 58.9 (34)  \\[0.2cm]
 Region~2 & 12: 01: 50.6& $0.11 (<0.27)$ & $0.33^{+0.09}_{-0.06}$ & $0.034 (<0.08)$ & $0.27^{+0.04}_{-0.04}$ \\[0.2cm]
$302.0\pm18.0$ &  -18: 52: 04.3 &  $<0.45$ & $2.8^{+1.3}_{-0.8}$ &  $0.51^{+0.44}_{-0.36}$ &$0.75^{+0.30}_{-0.18}$ \\[0.2cm] 
 & 137   & & 14.9 (13)  & & 11.2 (13) \\[0.2cm]
\enddata
\tablenotetext{a,c}{ Absorbing column density for the total spectrum
 in units of $10^{22}~\rm{cm^{-2}}$. Quoted errors are at the 90\% confidence level
for one interesting parameter.}
\tablenotetext{b,d}{ Temperature of the  thermal component in units of keV. Quoted errors are at the 90\% confidence level
for one interesting parameter.}
\end{deluxetable}
\begin{deluxetable}{lccccccccccc}
\tablenum{4}
\tablecolumns{11}
\tablewidth{0pt}
\tablecaption{ Spectral Fits for Selected Regions (free Abundance)}
\tablehead{ \colhead{Region}&
\multicolumn{2}{c}{RS+(PO)} &  \multicolumn{2}{c}{RS+(RS)}   \\
\colhead{}  & \colhead{$\rm{N_{H,tot}}^{a}$ }  &\colhead{kT$^{b}_{1}$ } 
& \colhead{$\rm{N_{H,tot}}^{a}$ } & \colhead{kT$^{b}_{1}$ } & \colhead{Z$^{e}_{1}$ }  \\
\colhead{ }  &\colhead{$\rm{N_{H}}^{c}$ }  &\colhead{ $\Gamma$} &
\colhead{$\rm{N_{H}}^{c}$ } & \colhead{kT$^{d}_{2}$} & \colhead{Z$^{d}_{2}$} \\
\colhead{ }         & \colhead{Z$^{e}$} &\colhead{ \x2 (dof)}  &
 \colhead{ }  &\colhead{ \x2 (dof)}  &   \\
\colhead{ (1) } &\colhead{ (2)}
&\colhead{(3)} &\colhead{(4)} &\colhead{(5)} &\colhead{(6)}} 
\startdata
North Nucleus &  $0.14 (<0.23)$ & $0.63^{+0.07}_{-0.25}$ & $0.06 (<0.15)$ & $0.3^{+0.08}_{-0.06}$ & $0.17^{+1.89}_{-0.07}$\\[0.2cm]
&  $3.4 (<30.3)$ & $3.12^{+6.8}_{-3.18}$  & $0.46^{+0.45}_{-0.31}  $ &$0.7^{+0.12}_{-0.1}$ & $0.12^{+0.19}_{-0.19}$ \\[0.2cm]
& $0.10^{+0.02}_{-0.02}$ & 37.5 (28) & $0.46^{+0.45}_{-0.31}$& 32.5 (27)  \\[0.2cm]
South Nucleus & $0.10^{+0.02}_{-0.02}$ & $0.74^{+0.03}_{-0.07}$
& $0.034 (<0.1)$ &$0.35^{+0.06}_{-0.02}$ & $0.22 (>0.08)$ \\[0.2cm]
& $2.5 (<180.3)$ & $-0.45 (<6.5)$ &$0.17^{+0.13}_{-0.16}$ &$0.82^{+0.05}_{-0.05}$ & $0.23^{+0.27}_{-0.13}$ \\[0.2cm]
 & $0.14^{+0.05}_{-0.05}$ &135.1 (66) & & 99.3 (65)  \\[0.2cm]
Region~1 &  $0.034 (<0.108)$ &$0.37^{+0.14}_{-0.07}$  &$0.034 (<0.105)$ & $0.26^{+0.09}_{-0.07}$ & $0.09 (>0.02) $\\[0.2cm]
&  $1.24_{-0.49}^{+0.33}$ & $9.34 (>6.37)$  &$0.05 (<0.36)$ & $0.73^{+0.09}_{-0.09}$  & $0.29 (>0.09) $ \\[0.2cm]
&  $0.07_{-0.03}^{0.08}$  & 41.8 (33) & & 41.7 (32)  \\[0.2cm]
\enddata
\tablenotetext{a,c}{ Absorbing column density for the total spectrum
 in units of $10^{22}~\rm{cm^{-2}}$. Quoted errors are at the 90\% confidence level
for one interesting parameter.}
\tablenotetext{b,d}{ Temperature of the  thermal component in units of keV. Quoted errors are at the 90\% confidence level
for one interesting parameter.}
\end{deluxetable}

\begin{deluxetable}{lccccccccccc}
\tabletypesize{\scriptsize}
\tablenum{5}
\tablecolumns{11}
\tablewidth{0pt}
\tablecaption{ Spectral Fits for Selected Regions (Z=Z$\odot$)}
\tablehead{ \colhead{Region}&  \multicolumn{2}{c}{RS${_1}$ + (RS${_2}$+PO) } &
\multicolumn{2}{c}{RS${_1}$ + RS${_2}$ + (PO) }  
&  \multicolumn{2}{c}{RS${_1}$ + (RS${_2}$+RS${_3}$) } & \multicolumn{2}{c}{RS${_1}$ + RS${_2}$ + (RS${_3}$) } \\
\colhead{ }   & \colhead{$\rm{N_{H,tot}}^{a}$ }  &\colhead{kT$^{b}_{1}$ } 
& \colhead{$\rm{N_{H,tot}}^{a}$ } & \colhead{kT$^{b}_{1}$ } 
& \colhead{$\rm{N_{H,tot}}^{a}$ } & \colhead{kT$^{b}_{1}$ } 
& \colhead{$\rm{N_{H,tot}}^{a}$ } & \colhead{kT$^{b}_{1}$ } \\
\colhead{ }  & \colhead{$\rm{N_{H}}^{c}$ } & \colhead{kT$^{d}_{2}$} &
\colhead{$\rm{N_{H}}^{c}$ } & \colhead{kT$^{b}_{2}$}& 
\colhead{$\rm{N_{H}}^{c}$ } & \colhead{kT$^{d}_{2}$} & 
\colhead{$\rm{N_{H}}^{c}$ } & \colhead{kT$^{b}_{2}$}\\
\colhead{ }      &  
 \colhead{ $\Gamma$}  & \colhead{ \x2 (dof)} & 
\colhead{ $\Gamma$} &  \colhead{ \x2 (dof)} &
\colhead{ $kT^{d}_{3}$}  & \colhead{ \x2 (dof)} & 
\colhead{ $kT^{d}_{3}$} &  \colhead{ \x2 (dof)} \\
\colhead{ (1) } &\colhead{ (2)}
&\colhead{(3)} &\colhead{(4)} &\colhead{(5)} & \colhead{ (6) }  &\colhead{ (7)}
&\colhead{(8)} &\colhead{(9)} }
\startdata
North Nucleus & $0.034 (<0.18)$        & $0.28^{+0.09}_{-0.06}$ & $0.034 (<0.21)$        & $0.28^{+0.10}_{-0.09}$ & $0.034 (<0.72)$        & $0.28^{+0.03}_{-0.04}$ & $0.034 (<0.075)$       & $0.29^{+0.05}_{-0.05}$ \\[0.2cm]
              & $0.24 (<0.69)$         & $0.70^{+0.11}_{-0.11}$ & $0.27 (<0.83)$         & $0.72^{+0.13}_{-0.11}$ & $0.73^{+0.19}_{-0.40}$ & $0.65^{+0.10}_{-0.07}$ & $0.68^{+0.20}_{-0.37}$ & $8.1   (>2.7)  $       \\[0.2cm]
              & $2.87^{+1.08}_{-0.64}$ & 30.0 (27)              & $2.96^{+1.44}_{-0.59}$ &  30.6 (27)             & $16.2  (>1.9)$         & 33.2 (27)              & $0.66^{+0.09}_{-0.09}$ &  30.7 (27)             \\[0.2cm]
South Nucleus & $0.034 (<0.10)$        & $0.35^{+0.06}_{-0.03}$ & $0.034 (<0.15)$        & $0.37^{+0.06}_{-0.03}$ & $0.034 (<0.072)$       & $0.36^{+0.05}_{-0.04}$ & $0.053 (<0.073)$       & $0.36^{+0.05}_{-0.03}$ \\[0.2cm]
              & $0.15^{+0.11}_{-0.12}$ & $0.82^{+0.06}_{-0.04}$ & $0.33 (<0.6)$          & $0.85^{+0.06}_{-0.05}$ & $0.03 (<0.08)$         & $0.85^{+0.05}_{-0.04}$ & $  <0.09$              & $0.85^{+0.11}_{-0.03}$\\[0.2cm]
              & $2.9^{+0.8}_{-0.5}$    & 99.7 (65)              & $3.47^{+1.24}_{-1.09}$ & 101.2 (65)             & $2.9^{+2.0}_{-0.08}$   & 112.8 (65)             & $2.94^{+1.92}_{-0.82}$ & 113.1 (65)             \\[0.2cm]
Region~1      & $0.034 (<0.13)$        & $0.25^{+0.07}_{-0.06}$ & $0.06 (<0.14)$         & $0.25^{+0.08}_{-0.09}$ & $0.21^{+0.12}_{-0.13}$ & $0.07^{+0.03}_{-0.03}$ & $0.23^{+0.10}_{-0.10}$ & $0.06^{+0.05}_{-0.02}$ \\[0.2cm]
              & $0.05 (<0.19)$         & $0.74^{+0.08}_{-0.09}$ & $<0.32$                & $0.75^{+0.08}_{-0.08}$ & $0.05 (<0.3)$          & $0.28^{+0.04}_{-0.05}$ & $0.032  (<0.29)$       & $0.28^{+0.04}_{-0.05}$   \\[0.2cm]
              &  $3.53^{+4.0}_{-0.9}$  & 42.4 (32)              & $3.37^{+1.58}_{-0.87}$ & 42.6 (32)              & $0.79^{+0.11}_{-0.08}$ & 42.8 (32)              & $0.79^{+0.10}_{-0.11}$ & 42.9 (32)              \\[0.2cm]
\enddata
\tablenotetext{a,c}{ Absorbing column density for the total spectrum
(a) and the hard component (c) in units of $10^{22}~\rm{cm^{-2}}$. Quoted errors are at the 90\% confidence level
for one interesting parameter.}
\tablenotetext{b,d}{ Temperature of the unabsorbed (b) and absorbed
(d) thermal component in units of keV. Quoted errors are at the 90\% confidence level
for one interesting parameter.}
\end{deluxetable}

\begin{deluxetable}{lccccccccccc}
\tabletypesize{\scriptsize}
\tablecolumns{11}
\tablenum{6}
\tablewidth{0pt}
\tablecaption{ Intensity of Spectral Components (Z=Z$\odot$)}
\tablehead{ \colhead{Region}&  \multicolumn{2}{c}{RS${_1}$ + (RS${_2}$+PO) } &  \multicolumn{2}{c}{RS${_1}$ + RS${_2}$ +(PO) } &  \multicolumn{2}{c}{RS${_1}$ + (RS${_2}$+RS${_3}$) } &  \multicolumn{2}{c}{RS${_1}$ + RS${_2}$ + (RS${_3}$) }   \\
\colhead{ }     & \colhead{EM$_{1}^{a}$ ($\rm{L_{X}}$)} & \colhead{EM$_{2}^{b}$ ($\rm{L_{X}}$) } & \colhead{EM$_{1}^{a}$ ($\rm{L_{X}}$) } & \colhead{EM$_{2}^{b}$ ($\rm{L_{X}}$) } & 
\colhead{EM$_{1}^{a}$ ($\rm{L_{X}}$)} & \colhead{EM$_{2}^{b}$ ($\rm{L_{X}}$) } & \colhead{EM$_{1}^{a}$ ($\rm{L_{X}}$) } & \colhead{EM$_{2}^{b}$ ($\rm{L_{X}}$) } \\
\colhead{ }     & \colhead{Norm.$^{c}$ ($\rm{L_{X}}$) } &  \colhead{ } & \colhead{Norm.$^{c}$ ($\rm{L_{X}}$)} &  \colhead{ }     & 
 \colhead{EM$_{3}^{d}$ ($\rm{L_{X}}$) } &  \colhead{ } & \colhead{EM$_{3}^{d}$ ($\rm{L_{X}}$)} &  \colhead{ }  \\
 \colhead{ (1) } &\colhead{ (2)}
&\colhead{(3)} &\colhead{(4)} &\colhead{(5)} & \colhead{ (6) }  &\colhead{ (7)}
&\colhead{(8)} &\colhead{(9)} }
\startdata
North Nucleus & 2.9 (0.38) & 5.3 (0.89) &   2.3 (0.30)   & 2.6 (0.43)   &     4.62 (0.60) & 28.8 (4.96)  & 4.03 (0.52) &4.87 (0.47)    \\[0.2cm]  
              & 7.9 (4.53) &                  &   9.6 (6.28)   &                    &     4.21 (0.40) &                    & 22.8 (3.93)                       \\[0.2cm]  
South Nucleus & 9.7 (1.32) &26.1 (3.70) &   8.9 (1.24)   & 17.1 (2.34)  &     10.1 (1.39) & 21.2 (2.88)  & 10.9 (1.52) &20.3 (2.74)    \\[0.2cm]  
              &11.9 (7.38) &                  &  23.7 (33.0)   &                    &     16.5 (1.27) &                    & 16.6 (1.28)                       \\[0.2cm]  
Region~1      & 4.6 (0.60) & 5.8 (0.92) &   5.6 (0.73)   & 5.4  (0.86)  &   350.2 (338.7)& 17.4 (2.25)  & 95.6 (96.6)&19.7 (2.54)    \\[0.2cm]  
              & 2.6 (3.95) &                  &   2.4 (3.04)   &                    &     8.4 (1.26)  &                    & 7.95 (1.19) &                     \\[0.2cm]  
Region~2$^{e}$& 3.4 (0.44) & 6.2 (0.90) &                      &                    &   4.0 (0.53)    &                    &                  &                      \\[0.2cm]  
              &                  &                  &                      &                    &   2.4 (1.27)    &                    &                  &                      \\[0.2cm]  
\enddata
\tablenotetext{a,b}{ Emission Measure 
for the low (a) and high (b)
temperature component in units of $10^{8}\times(4\pi
D^{2})~\rm{cm^{-3}}$  where D is the distance. 
In parentheses, absorption-corrected  luminosity in the
0.1-10.0keV band, in units of $10^{39}$~\ergs }
\tablenotetext{c}{ Normalization 
of the PO component (at 1~keV) in units of
$\rm{10^{-6}~photons/keV/cm^{2}/s}$. 
In parentheses, absorption-corrected  luminosity in the
0.1-10.0keV band, in units of $10^{39}$~\ergs }
\tablenotetext{d}{  Emission Measure (and luminosity) for the third thermal
component. The units are the same as in the first two components. }
\tablenotetext{e}{ The parameters for Region~2 correspond to the  RS+(PO) and RS+(RS) models instead of the RS+(RS+PO) and RS+(RS+RS)  models respectively. }
\end{deluxetable}

\begin{deluxetable}{lcccccc}
\tablecolumns{7}
\tablenum{7}
\tablewidth{0pt}
\tablecaption{Estimated ISM Parameters}
\tablehead{\colhead{Region}& \colhead{n} & \colhead{$E_{th}$} & \colhead{$\tau_c$}& \colhead{$M_{ISM}$}& \colhead{$p$} &\colhead{SNR}\\
& \colhead{($10^{-2}$ cm$^{-3}$)}& \colhead{($10^{54}$ ergs)}& \colhead{($10^7$ yrs)} & \colhead{($10^6 M_{\odot}$)} &\colhead{10$^{-11}$dyn~cm$^{-2}$} &\colhead{($10^{-3}$ yrs$^{-1}$)}}
\startdata
North Nucleus & 3.3$_1$, 4.5$_2$ & 0.5$_1$, 1.6$_2$ & 5$_1$, 6$_2$ & 0.3$_1$, 0.4$_2$ & 3.1$_1$, 10.7$_2$& 0.2\\
& (1.5$_1$, 2.0$_2$)$^a$ & (1.2$_1$, 3.6$_2$) & (10$_1$, 13$_2$) & (0.7$_1$, 0.9$_2$) & (1.4$_1$, 4.8$_2$) & (0.3)\\[0.20cm]
South Nucleus & 1.9$_1$, 3.2$_2$ & 3.0$_1$, 14$_2$ & 7.7$_1$, 12$_2$ & 1.7$_1$, 3$_2$ & 2.3$_1$, 8.9$_2$ & 1.2\\
& (0.9$_1$, 1.4$_2$) & (7.1$_1$, 30$_2$) & (18$_1$, 27$_2$) & (4$_1$, 6.4$_2$) & (1.1$_1$, 3.9$_2$) & (3.0)\\[0.20cm]
Region 1 & 2.2$_1$, 2.5$_2$ & 1.0$_1$, 3.5$_2$ & 5.4$_1$, 13$_2$ & 0.7$_1$, 0.8$_2$ & 1.9$_1$, 6.3$_2$ & 0.3\\
& (1.0$_1$, 1.1$_2$) & (2.1$_1$, 7.7$_2$) & (12$_1$, 28$_2$) & (1.6$_1$, 1.7$_2$) & (0.8$_1$, 2.8$_2$) & (0.7)\\[0.20cm]
\enddata
\tablenotetext{1}{Lower temperature RS component}
\tablenotetext{2}{Higher temperature RS component}
\tablenotetext{a}{For an emitting cylinder of height 200~pc;
values in parentheses are for a 1000~pc height cylinder}
\end{deluxetable}

\end{document}